\documentclass[aps,preprintnumbers,superscriptaddress,twocolumn,amsmath,amssymb,floatfix,prl]{revtex4-1} 
\pdfoutput=1
\usepackage{graphicx}
\usepackage[usenames,dvipsnames]{color}
\usepackage[colorlinks,bookmarks=false,citecolor=NavyBlue,linkcolor=Red,urlcolor=blue]{hyperref}
\usepackage{float}

\usepackage[export]{adjustbox}

\usepackage[normalem]{ulem}

\usepackage{lmodern}
\usepackage{graphicx}
\usepackage{comment}
\usepackage[export]{adjustbox}
\usepackage{dcolumn}
\usepackage{bm}
\usepackage{hyperref}
\hypersetup{linktocpage,colorlinks,citecolor={blue},pdfdisplaydoctitle=true,pdfpagemode=UseOutlines,bookmarksnumbered=true}

\begin{document}

\title{Darcy law for yield stress fluid}
\author{Chen Liu}
\affiliation{FAST, CNRS, Univ. Paris-Sud, Universit\'{e} Paris-Saclay, 91405 Orsay, France}
\author{Andrea De Luca}
\affiliation{Theoretical Physics, Oxford University, Parks Road, Oxford OX1 3PU, United Kingdom}
\author{Alberto Rosso}
\affiliation{LPTMS, CNRS, Univ. Paris-Sud, Universit\'{e} Paris-Saclay, 91405 Orsay, France}
\author{Laurent Talon}
\affiliation{FAST, CNRS, Univ. Paris-Sud, Universit\'{e} Paris-Saclay, 91405 Orsay, France}

\newcommand{\andrea}[1]{{\color{red} #1}}

\begin{abstract}
{Predicting the flow of non-Newtonian fluids in porous structure is still a challenging issue due to the interplay betwen the microscopic disorder and the non-linear rheology. In this letter, we study the case of an yield stress fluid in a two-dimensional structure. Thanks to a performant optimization algorithm, we show that the system undergoes a continuous phase transition in the behavior of the flow controlled by the applied pressure drop. In analogy with the studies of the plastic depinning of vortex lattices in high-$T_c$ superconductors we characterize the nonlinearity of the flow curve and relate it to the change in the geometry of the open channels. In particular, close to the transition, an universal scale free distribution of the channel length is observed and explained theoretically via a mapping to the KPZ equation.  } 
\end{abstract}

\maketitle

Most of the water used for human consumption is stored in underground porous structures, called acquifers, where it is free to flow if a stress drop $P$ is applied.
 In 1852, H. Darcy \cite{darcy1856} showed that the mean debit, $Q$, namely the volume of fluid which passes per unit time measured in a region of size $L$, is proportional to the drop $P /L$: $Q= \left(\kappa/\eta\right) \frac{P}{ L}$. Here $\eta$ is the water viscosity and $\kappa$ is the permeability, which depends on the composition of the porous structure. The permeability can vary of many order of magnitude: from the quite large values of fractured rock or gravel to the extremely small permeability of clay. It is a macroscopic measure of the interplay between the liquid and the solid at the pore scale.

The Darcy's law is not restricted to underground water, but holds for oil, natural gas, and all Newtonian fluids embedded in porous structure. However it does not capture the behavior of many fluids currently injected in rocks for various applications.  In hydraulic fracturing, for example, cracks induced by high-pressure fluid injection allow the flow of gas and oil \cite{barbati16}. The fracking fluids are emulsions of water and sand or other proppants needed to keep the paths open.  Foams are used in the enhanced oil recovery (EOR) to avoid the viscous fingering instability~\cite{nittmann1985fractal}. Complex fluids are also employed in soil consolidation by cement injection. All the aforementioned applications involve yield stress fluid, namely liquids that are able to flow only above 
a finite yield stress, $\tau_y$. Thus, it is an important question to understand how yield stress liquids flow in the ground and in general in porous materials \cite{entov67,al-fariss87,talon13b,chevalier13,chevalier15a,chevalier15c}.
Recent studies have shown that, due to structural disorder in the  material, flow occurs only above  a critical pressure drop $P_0$ and it is characterized by a phase separation with regions that are easier to flow than others \cite{talon13a,chevalier15a,chevalier15c}. In this regime, the flow curve is non linear with $Q \propto (P - P_0)^\beta$ with $\beta>1$. 
At higher pressure, linearity is recovered and the flow invades homogeneously the material. 

\begin{figure}[h]
	\begin{center}
		\begin{minipage}{0.46\hsize}
			\includegraphics[width =\hsize]{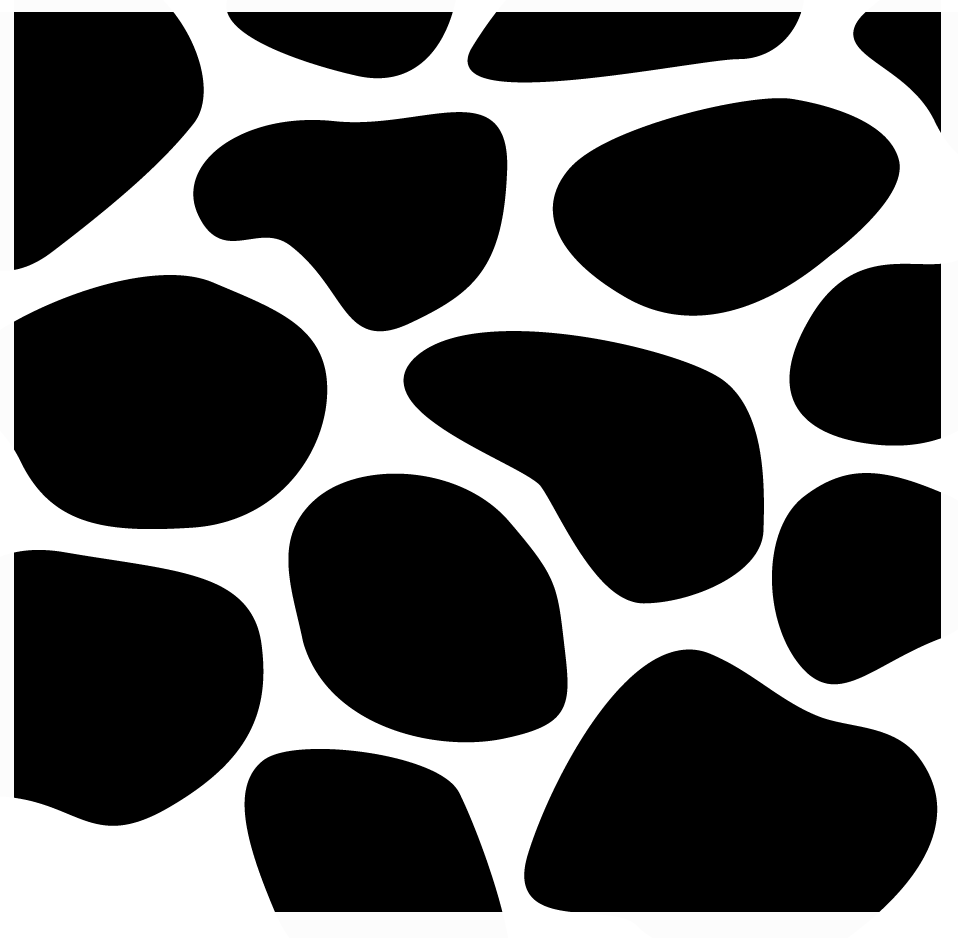}
		\end{minipage}
		\begin{minipage}{0.49\hsize}
			\includegraphics[width =\hsize]{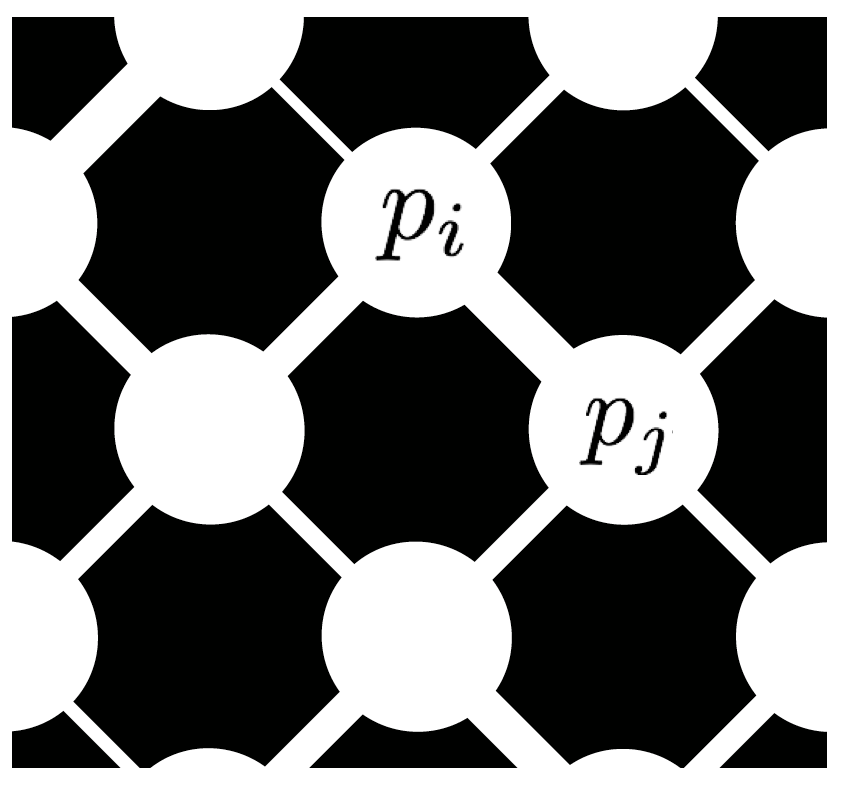}
		\end{minipage}
	\end{center}
	\caption{Sketch of porous media. Left: realistic porous medium in which the solid phase consists of an assembly of grains (in black).
		Right: model of a pore network in which large open pores are connected by straight tubes (throats) with random radius and unit length.} 
	\label{fig:model}
\end{figure}

Interestingly  similar behaviors are observed  in other disorder systems such as  vortex lattices in  high-Tc superconductor \cite{pardo1998observation,kolton1999hall}, skyrmions \cite{reichhardt2015collective} or 2D colloidal crystals \cite{reichhardt2002colloidal,pertsinidis2008statics}.
There, a \emph{plastic depinning} is observed  above a critical forcing, with vortices moving in preferential narrow channels and displaying a non-linear flux  with the applied force ($\beta>1$). At larger force, a smectic ordered phase is  observed  and the flux  becomes linear \cite{reichhardt2016depinning}.
In all these systems, a dynamical continuous transition separates an arrested phase from a flowing one.
In these conditions, universality and divergent correlation lengths are expected \cite{watson1996collective}, but never proved even though power-law behaviour has been reported in erosion models and experiments \cite{Aussillous201609023}.

In this letter, we provide a quantitative description  of yield stress fluids in a  stylized two-dimensional porous material.
Introducing a very performant optimization algorithm, we compute 
the  flow curve of large systems without approximations and observe three distinct regimes. In particular, for the plastic flow we find $Q \propto (P - P_0)^\beta$ with $\beta \simeq 2$.
When $P \rightarrow P_0$, the distribution $P_{\ell}$ of channels of length $\ell$ becomes scale free:
$P_{\ell} \propto 1/\ell$.
A mapping with the model of directed polymer in random media allows to show that this result is universal and belongs to the KPZ universality class \cite{kardar86}.

\paragraph{The model ---}
The full solution of the flow through a porous medium is computationally costly even for Newtonian fluids, since it requires to solve the Navier-Stokes equations coupled with the no-slip condition at the complex solid interface.  A significant simplification is provided by the pore network models \cite{fatt1956network} shown in Fig.~\ref{fig:model}. There the material is described by a lattice of large voids (the pores) connected by narrow cylindric tubes (the throats) of length $l$ and radius $r_0$.
 In the pores the pressure is assumed homogeneous, the flow occurs in the throats where it  can be computed even for a non-Newtonian fluids. In particular for a Bingham rheology \cite{bingham1922fluidity}
 the local debit in the throat connecting the pores $i$  and  $j$ writes 
\begin{equation}
\label{eq:Qsmall}
q_{ij}= \left\{ 
\begin{array}{cc}
\sigma_{ij} (\Delta p_{ij}-\tau_{ij}) \quad &\text{if:} \;  \Delta p_{ij}>\tau_{ij}\\
 0  \quad &\text{if:} \; |\Delta p_{ij}| < \tau_{ij} \\
\sigma_{ij}(\Delta p_{ij}+\tau_{ij}) \quad &\text{if:} \; \Delta p_{ij}<-\tau_{ij}\\
\end{array} 
\right .
\end{equation}
with the local pressure drop $\Delta p_{ij}=p_i-p_j$; the local hydraulic conductivity $\sigma_{ij} \sim r_0^4/l$  and the local pressure threshold $\tau_{ij} = \tau_y l/r_0$ . In principle both local conductivities and thresholds are random \cite{bird1987dynamics,talon14} but, for simplicity,  we set $\sigma_{ij}$ to unity and implement 
the disorder nature of  porous materials  only with the randomness of $\tau_{ij}>0$. Eq.~\eqref{eq:Qsmall} should be combined with the Kirchhoff's conservation of the flow at each node$\sum_{j\in n(i)}q_{ij}=0$, 
where the sum runs over the set $n(i)$ of neighbours of the node $i$.  This conservation holds for all nodes except the inlet node where the fluid is injected at pressure $P$ and the outlet node where the fluid is evacuated at zero pressure. 
Given the total pressure drop $P$ and the configuration of random thresholds, Eq.~\eqref{eq:Qsmall} 
together with Kirchhoff's condition is closed, but very difficult to solve due to the nonlinearity of the flow rate function of Eq.~(\ref{eq:Qsmall}).

\begin{figure*}[th]
	\begin{center}
		\begin{minipage}{1\hsize}
			\includegraphics[width =1. \hsize]{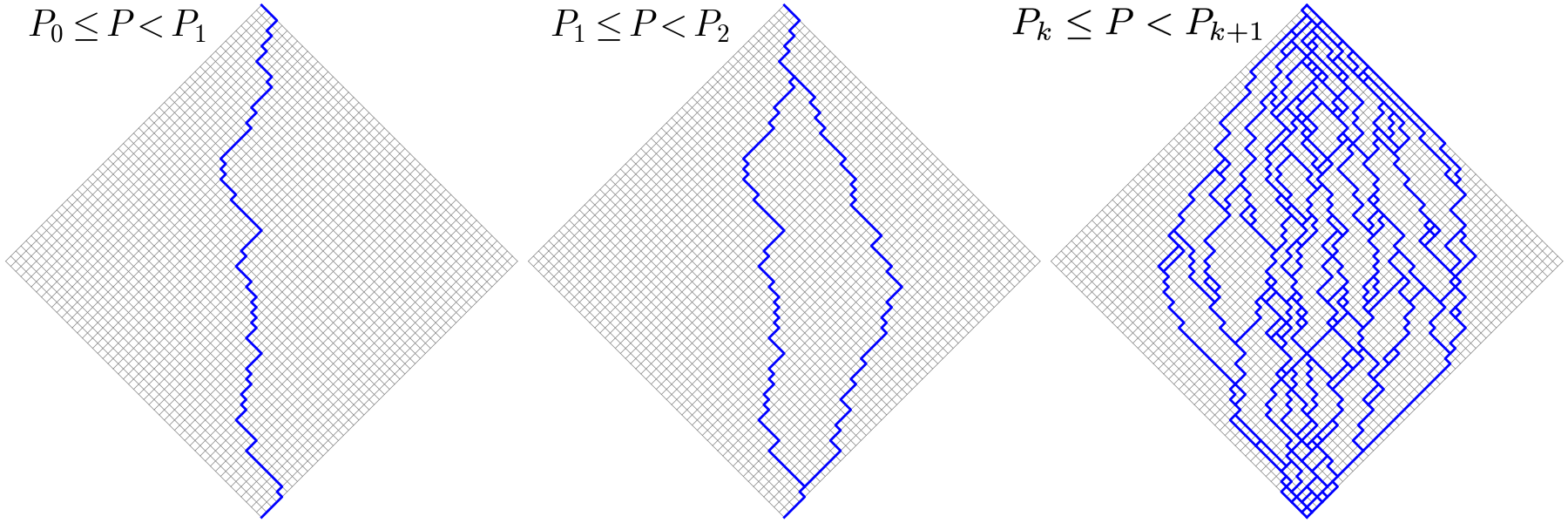}
		\end{minipage}
	\end{center}
	\caption{The flowing path network at different applied pressures for a system of size $L=100$. } 
	\label{fig:FlowFieldRegimes}
\end{figure*}

To resolve this task, we introduce two numerical methods that  apply to any lattice type.
For simplicity, we consider the 2D square lattice of Fig.~\ref{fig:FlowFieldRegimes} with a single inlet and outlet node. A directed path connecting the two nodes has thus size $L$ and we set $l=1$.

 For the first method, we note that the solution of the system is equivalent to the minimization of the functional  $F(\{p_{i}\},P)\equiv \frac{1}{2}\sum_{ij} \left[q_{ij}\left( p_i,p_j \right)\right]^2 $
with the constrain of a pressure drop $P$ between  the inlet and the outlet node.
Hence we can use numerical gradient descent to minimize the functional with respect to the local pressure field and find the solution up to certain numerical precision.
 However the computational cost becomes extremely high  in the critical region, when  $P$ gets close to $P_0$.
 
For this, we developed a second method based on the observation that,  for a given pressure drop, the flow occurs only in the set of open throats, named $\mathcal{L}(P)$.
Once  $\mathcal{L}(P)$ is known, the solution of the local pressure becomes linear  and  can be written as $p_i = a_i P +b_i$,
where the coefficients  $a_i, b_i$ depend  on $\mathcal{L}(P)$ and their expressions are given in the SI. 
 The set of open throats, $\mathcal{L}(P)$, is determined iteratively starting from the minimal pressure $P_0$ needed to open the first channel  connecting  the inlet and the outlet pores. $P_0$ is obtained by minimizing:
\begin{equation}\label{eq:minpress1}
P_0= \min_{C \in \mathcal{C}_{\text{in-out}}} \sum_{(ij) \in C } \tau_{ij}.
\end{equation}
where $\mathcal{C}_{\text{in-out}}$ represents the set of all directed paths connecting the inlet and outlet nodes.
The channel $C_0$ corresponds to the path that realizes the minimum so that $\mathcal{L}(P _0) =C_0$.
 For slightly larger pressure, the flow remains restricted  to this channel and thus $\mathcal{L}(P )=\mathcal{L}(P _0)$. Increasing the pressure, $\mathcal{L}(P )$  is enlarged as new channels will open.
  The changes of $\mathcal{L}(P )$  are the manifestation of the nonlinearity  of the problem and, for a given realization of the thresholds,  occur at precise  pressure values $P_0 < P_1 < P_2 < \ldots $ as shown in Fig.~\ref{fig:FlowFieldRegimes}.
  In order to determine $P_k$, and the corresponding  $\mathcal{L}(P_{k})$, knowing $\mathcal{L}(P_{k-1})$, we consider the ensemble $\mathcal{C}_{mn}$ of  all paths that connect a pair of nodes $n,m$ belonging to $\mathcal{L}(P_{k-1})$ and that avoid any other intersection with  $\mathcal{L}(P_{k-1})$.
   The optimal path among them  $\mathcal{C}_{mn}$ has a threshold
\begin{equation}\label{eq:minpress2}
E_{mn}= \min_{C \in \mathcal{C}_{mn}} \sum_{(ij) \in C } \tau_{ij}.
\end{equation}

For a given $P>P_{k-1}$, if the threshold $E_{mn}$ is larger than the corresponding pressure drop $\Delta p_{mn}(P)$ for all pairs of nodes $(m,n)$, then no new channels appear and  $\mathcal{L}(P) = \mathcal{L}(P_{k-1})$.
Expressing $p_m$ and $p_n$ in terms of $a_m$, $b_m$ and $a_n$, $b_n$ respectively, the pressure $P_k$ is then defined as:
\begin{equation}\label{eq:minpress3}
	P_{k}=  \min_{(m,n) \in \mathcal{L}(P_{k-1}) } \frac{E_{mn} -(b_m-b_n)}{a_m-a_n}.
\end{equation}


The  minimizations of Eq. \eqref{eq:minpress1} and Eq. \eqref{eq:minpress2} are performed using the Dijkstra optimization algorithm \cite{dijkstra1959note} which is quadratic in the path length. In principle the channel can be non-directed, but in practice the statistics is dominated by the directed ones \cite{hansen04,talon13a} and for simplicity we restrict our analysis to them. 

Once the local pressures $p_i$ are known, the total flow $Q$ is given by the outgoing flow from the inlet node.  
In particular in our model,  the flow curve reads $Q(P) = \kappa_k (P-P_{k-1})/L + Q(P_{k-1}) \quad \text{if}\; P_{k-1}<P<P_{k}$,  
where  $\kappa_k$  is the permeability of the set $\mathcal{L}(P_k)$ and is independent on the local thresholds (see SI).

\begin{figure}[th]
	\begin{center}
		\begin{minipage}{0.49\hsize}
			\includegraphics[width=\hsize,trim = 50 0 10 0, clip=false]{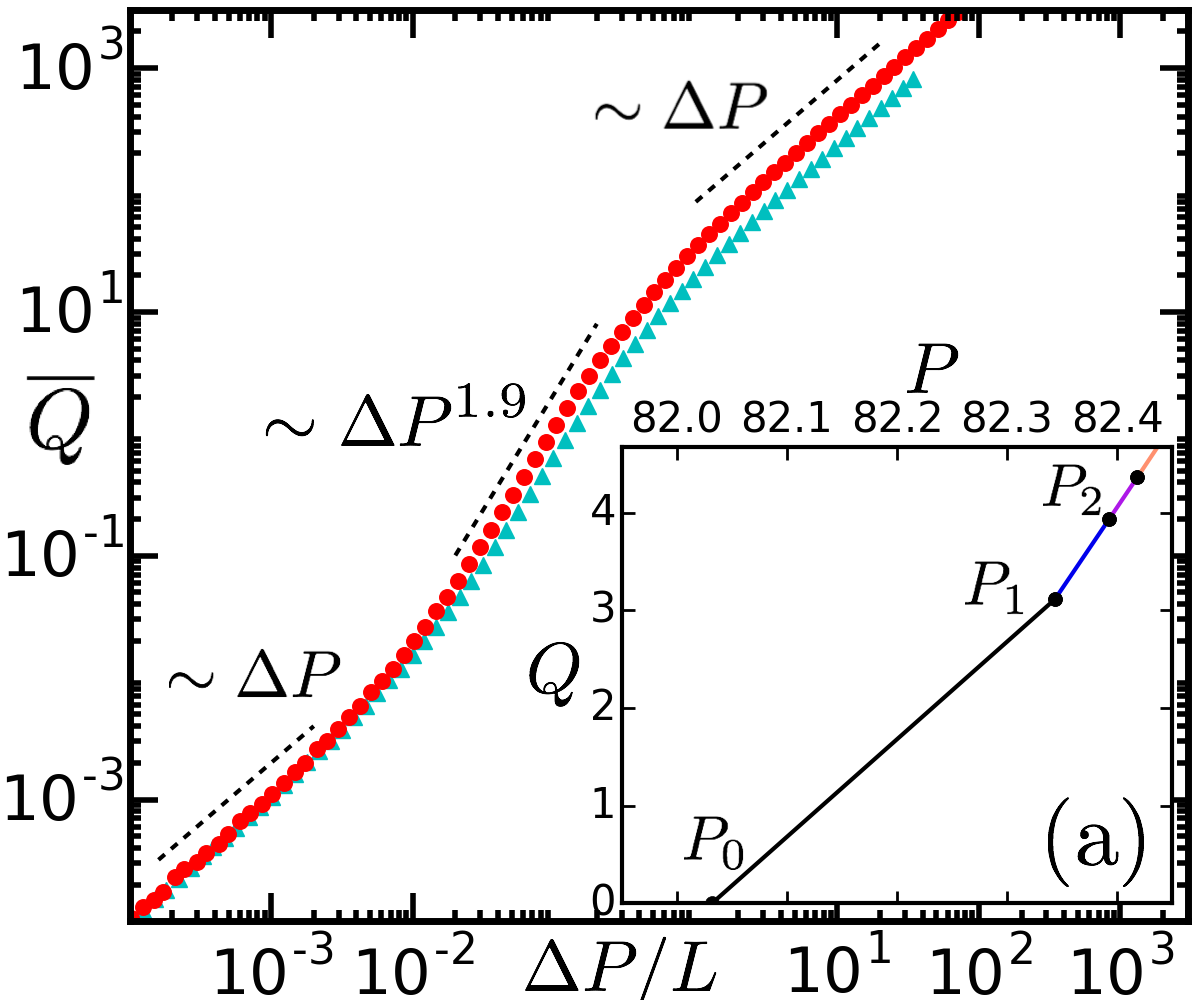}
		\end{minipage}
		\begin{minipage}{0.49\hsize}
			\includegraphics[width=\hsize,trim = 10 0 50 0, clip=false]{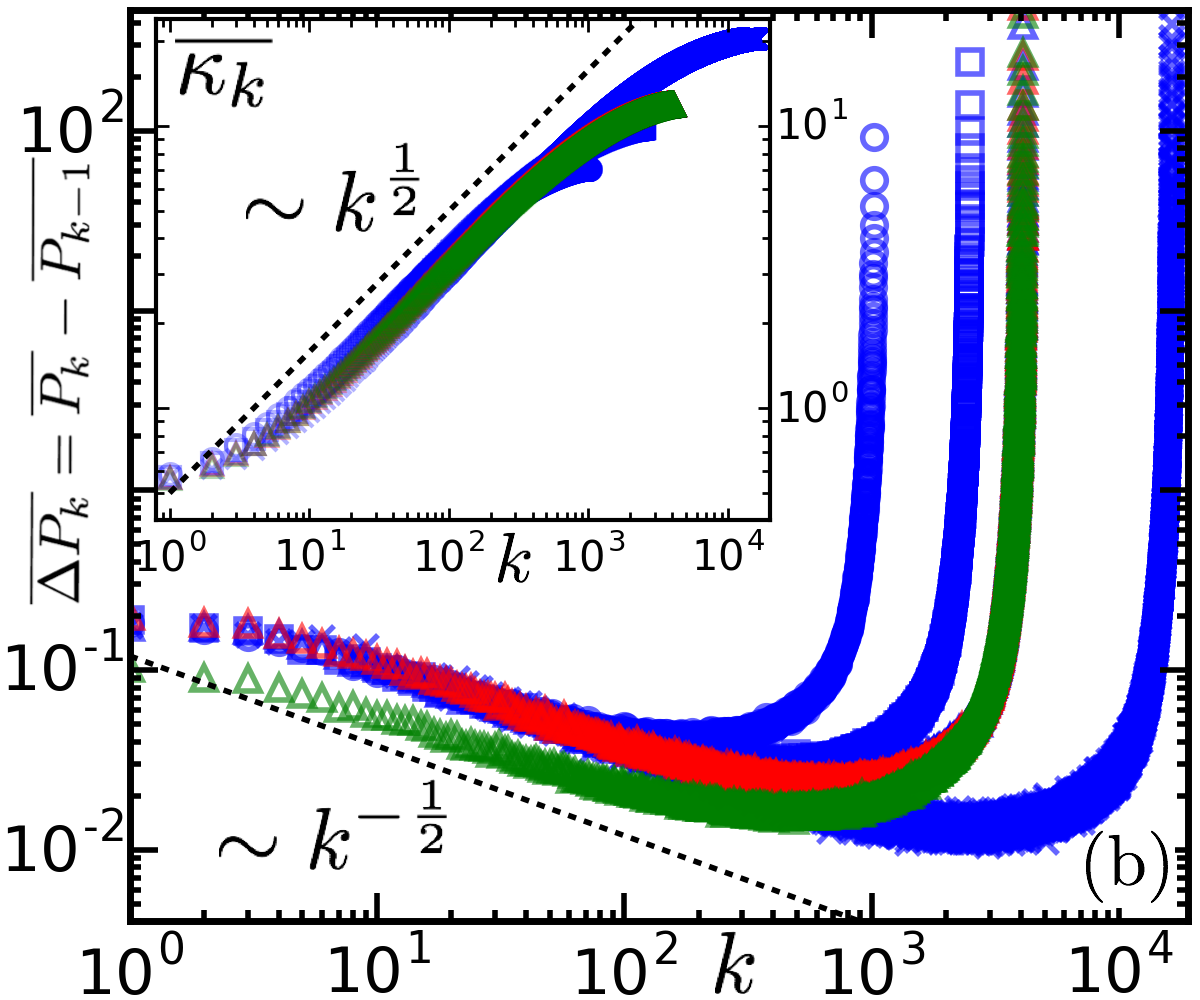}
		\end{minipage}
		\begin{minipage}{0.49\hsize}
			\includegraphics[width=\hsize,trim = 50 0 10 0, clip=false]{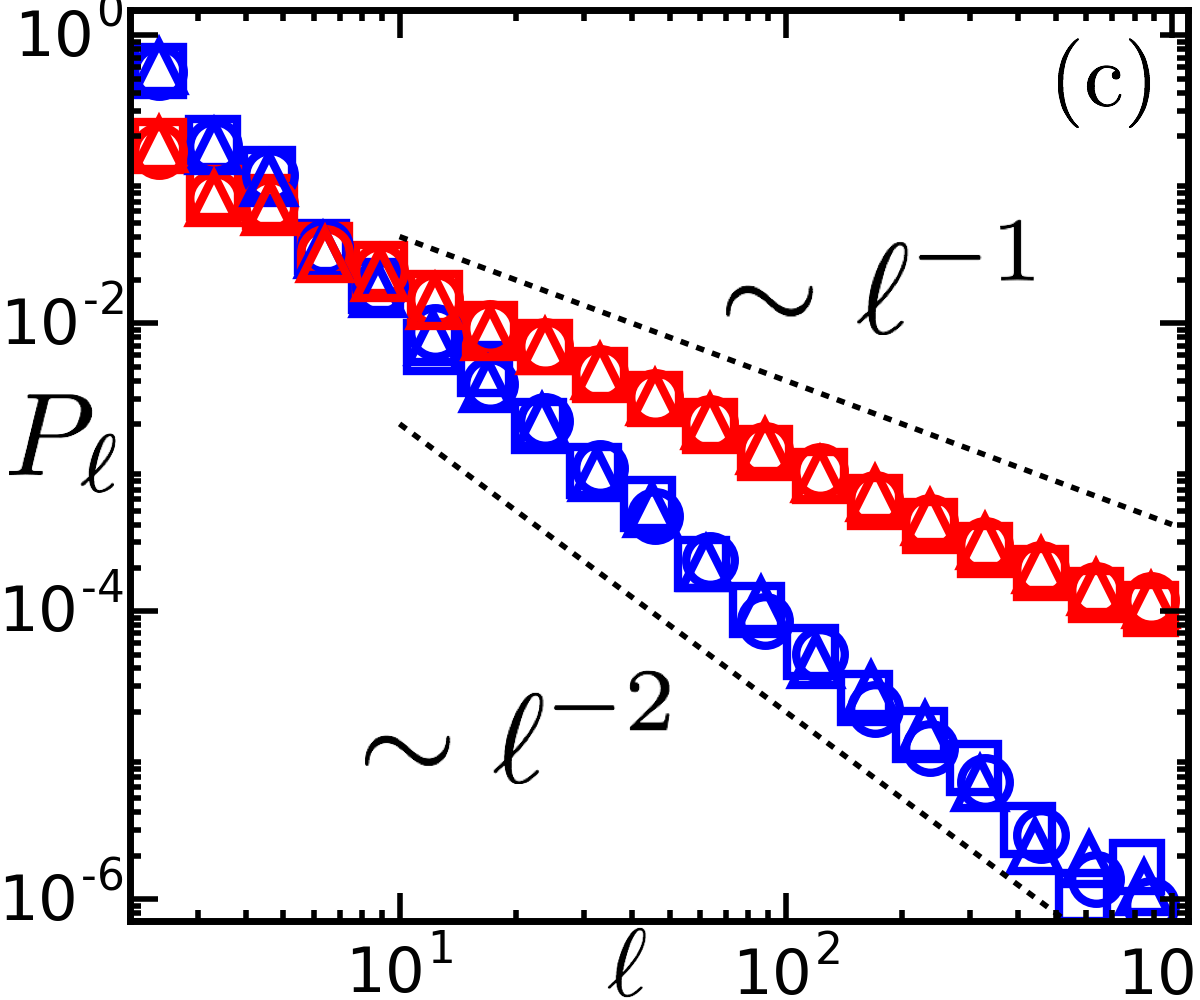}
		\end{minipage}
		\begin{minipage}{0.49\hsize}
            \includegraphics[width=\hsize,trim = 10 0 50 0, clip=false]{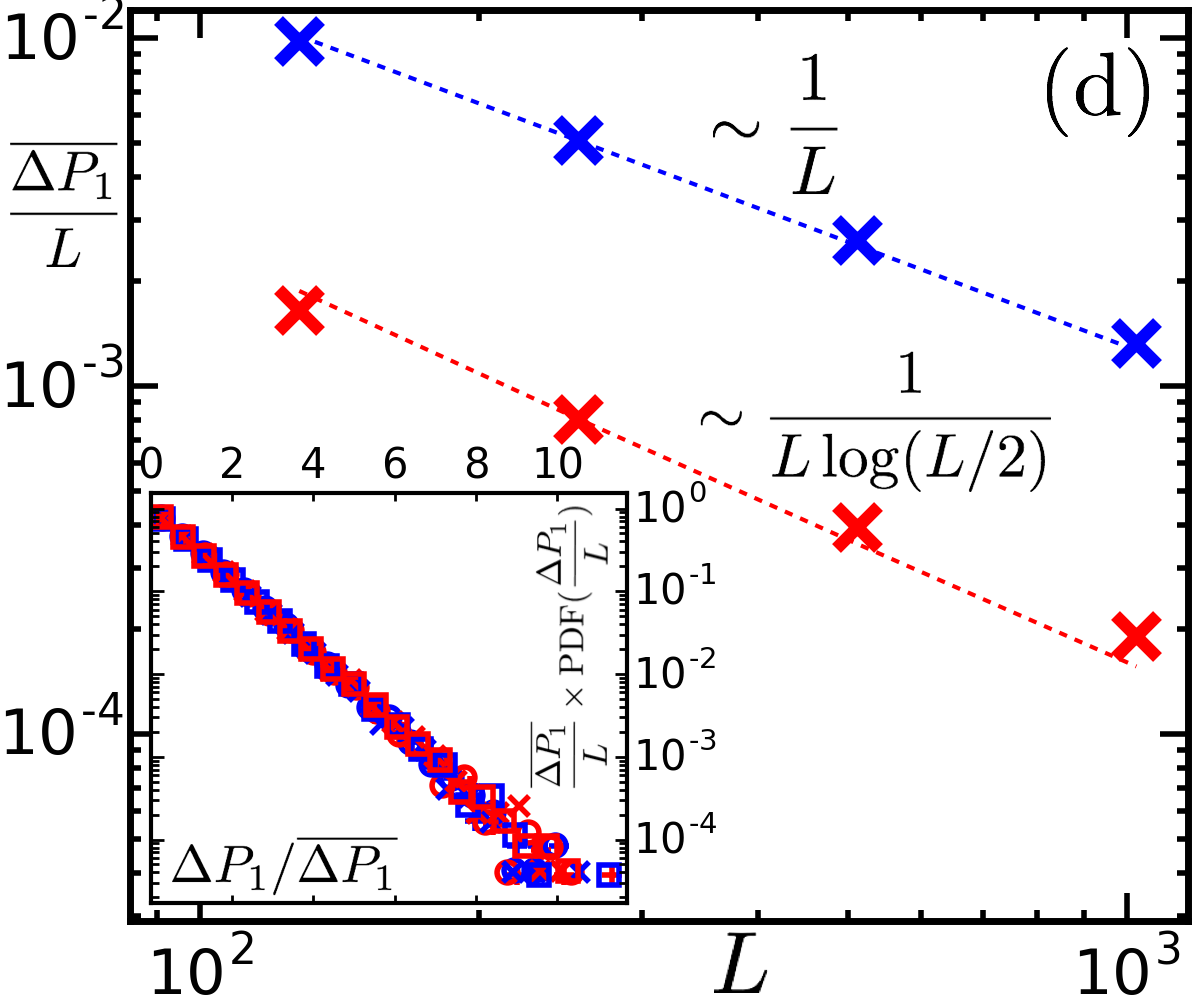}
		\end{minipage}
	\end{center}
	\caption{(a) The mean flow curve $\overline{Q}$ for a given $\Delta P= P-P_0$ averaged over more than $200$ realizations. The thresholds $\tau_{ij}$ are uniformly distributed in $[1.52, 2.08]$. Circles and triangles correspond to $L=64$ and $L=128$ respectively. Inset: the flow curve of a single realization($L=50$). (b) The averaged pressure increments $\overline{\Delta P_k}$ as function of the path number $k$. Inset: the averaged permeability $\overline{\kappa_k}$ as function of the path number $k$. Circles, squares, triangles and crosses correspond to different system sizes $L=64, 100, 128, 256$; Blue, red, green correspond to different types of disorder respectively: uniform, Gaussian, exponential. In (c) and (d), red symbols show the result for the present problem,  blue ones for the case where the factor $1/\ell$ is neglected in the minimization of Eq.~\eqref{eq:P1}.
	 (c) The PDF of lengths of the second open paths. Circles, squares and triangles correspond to different type of disorder:  uniform, Gaussian and exponential respectively. (d) The mean gap versus the system size (uniform distribution). Dashed lines represent the theoretical predictions. Inset: the PDFs of gaps for different system sizes renormalized by their mean values which follow a clear exponential distribution.} 
	\label{fig:central_results}
\end{figure}
%

\begin{figure}[th]
	\includegraphics[width =\hsize]{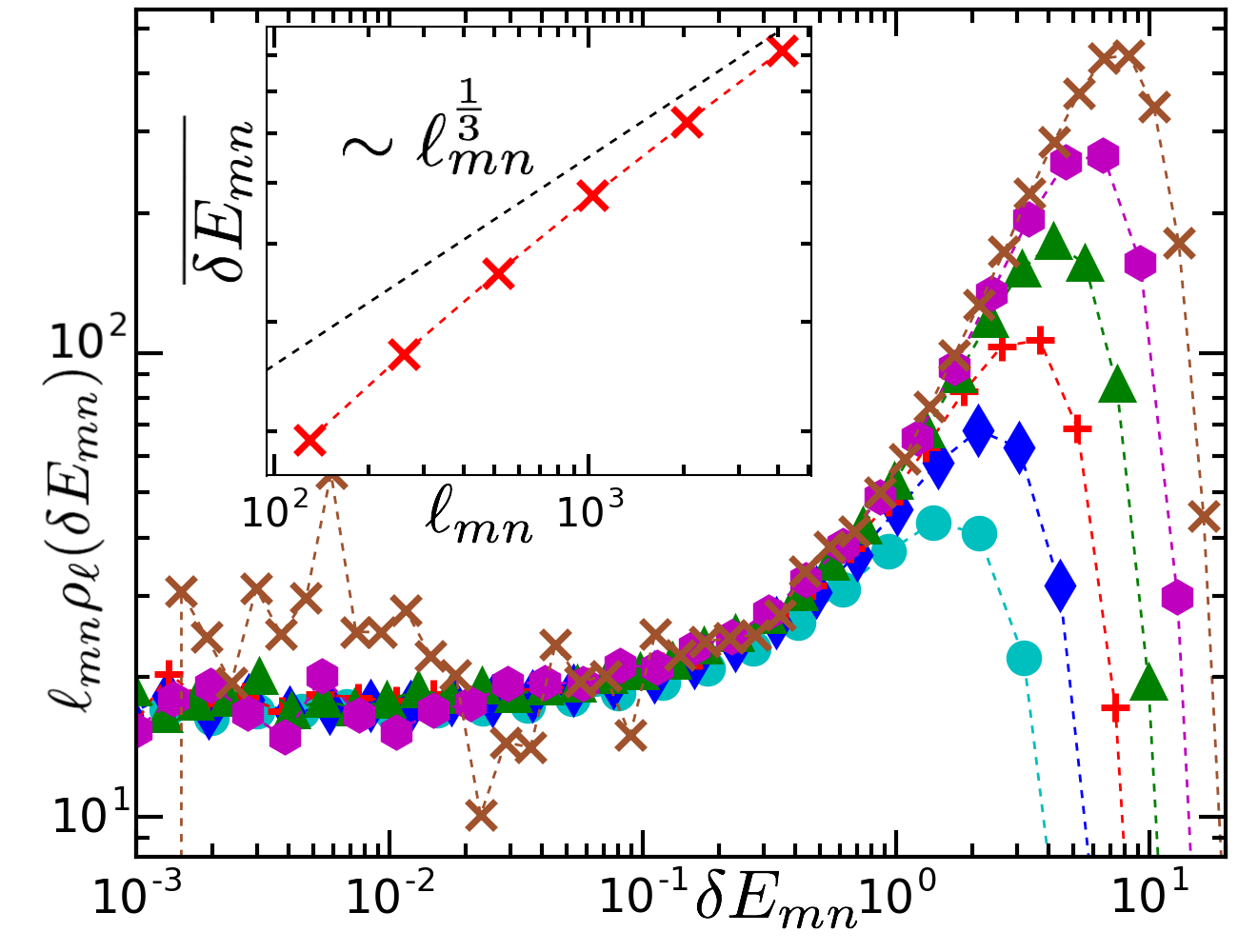}
	\caption{The rescaled PDFs of $\delta E_{mn}$ for different lengths $\ell_{mn}=128, 256, \ldots,4096$ from circles to crosses. The data collapse when $\delta E_{mn}\rightarrow 0$ indicates $\rho_{\ell}(\delta E\rightarrow 0 ) \sim 1/\ell$. Inset: the mean value $\overline{\delta E_{mn}}$ as function of the length $\ell_{mn}$. The dashed line represents $\ell^{1/3}$.}
	\label{fig:gap_stats} 
\end{figure}

\paragraph{Results ---}  
The flow curve (Fig.~\ref{fig:central_results}.a) shows two linear regimes at low and high pressures:  close to $P_0$ the permeability is $\kappa_1=1$ while, at very large pressure, it converges to the Newtonian value ($\tau_{ij}=0$):  $\kappa_{\infty} \sim \pi  L/(4 \log L)$ with a total flow: $Q(P) =(\kappa_\infty  P -\sum_{ij} \tau_{ij})/L$.

The regime at intermediate pressure is extremely interesting: it is characterized by a complex geometry of open channels (Fig.~\ref{fig:FlowFieldRegimes}.c) and by a non-linear growth of the flow rate (Fig.~\ref{fig:central_results}.a). A power law behavior emerges with an exponent $\beta$ close to $2$. The origin of this nonlinearity comes from the opening of new channels which increases the permeability (as shown in the inset of Fig.~\ref{fig:central_results}.a).
Indeed, in Fig.~\ref{fig:central_results}.b we study the sequence of pressure gap $\Delta P_k\equiv P_k-P_{k-1}$ and the permeability $\kappa _k$ as a function of the number of open paths $k$. Independently on the type of disorder, a clear power-law behavior is observed and we can conclude that the number of paths grows quadratically with $P-P_0$ while the flow rate grows as $Q =\frac{1}{L}\sum _{k'=1}^k \kappa _{k'} \Delta P_{k'} \sim k$, which is consistent with $\beta=2$. 
The value of the exponent seems then to be independent of the  type of disorder (see also  SI) and is also in agreement with the one found by solving the full Bingham rheology problem in  realistic porous structures (\emph{e.g.} random beads packing)  \cite{talon13a,chevalier15a}. This suggests that the macroscopic rheology might be universal and that the system approaching $P_0$ is undergoing a dynamical second-order phase transition. 
To assess this idea,  we then search for fingerprints of criticality and scale-free behaviours.

In particular, we can provide a much deeper understanding on how the nonlinearity of the flow is approached from small $P$. In the inset of Fig.~\ref{fig:central_results}.a, we present the flow curve for a single realization for $P \gtrsim P_0$. The exact linearity terminates at $P = P_1$ when a second path opens. In this  case, Eq.~\eqref{eq:minpress3} can be simplified to
\begin{equation}
\label{eq:P1}
P_{1}=  P_0 + L \min_{(m,n) \in \mathcal{L}(P_{0}) } \left [\frac{\delta E_{mn}}{\ell_{mn}} \right].
\end{equation}
where $\ell_{mn}$ is the distance between $m$ and $n$ and $\delta E_{mn} = E_{mn} - E_{mn}^0$, with $E_{mn}^0$ the threshold along $C_0$ between $n$ and $m$.
The presence of the second path $C_1$ induces a change on the permeability that depends on the length $\ell$ of $C_1$. In particular, one has $\kappa_1 = \frac{L}{L - \ell /2}$.
Note that, depending on the size of the re-organization $\ell$, the permeability can increase up to a factor $2$ for $\ell \sim L$.
Remarkably as shown in Fig.~\ref{fig:central_results}, the statistics of the re-organization size is a power law decaying as $P(\ell)\propto 1/\ell$, which characterize a scale-free behaviour.
We can explain this result by using a mapping with the directed polymer in random media and their connection with the KPZ universality class~\cite{kardar86,kardar87,roux87}.

\paragraph{Directed polymers and KPZ. ---}
The minimization problem in Eq.~\eqref{eq:minpress1} is equivalent to finding the ground state of a directed polymer (DP) of length $L$ in $1+1$ random medium and the pressure $P_0$ corresponds to its energy \cite{kardar86}. DP is a well-known model that belongs to the KPZ universality class and the sample-to-sample fluctuations of the ground state energy have been largely studied in the literature: they display an anomalous scaling $\propto L^{1/3}$~\cite{kardar86,kardar87,huse1985pinning} with a Tracy-Widom distribution~\cite{johansson2000shape}. 
The second path that opens at $P_1$ can be seen as an excited state, on which much less is known. It is natural to expect that the  \emph{constraint} of avoiding the ground state along the segment delimited by $(n,m)$ produces a gap value that grows  with the re-organization length $\ell_{mn}$, In particular,
one expects that $\delta E_{mn}$ grows as $ \ell_{mn}^{1/3}$~\cite{hwa1994anomalous}. If this scaling form is plugged in Eq.~\eqref{eq:P1}, one obtains the saturation 
of the typical size of the re-organization $\ell \simeq L$ and a gap growth  $P_1 - P_0 \propto L^{1/3}$.
 Note that if in Eq.~\eqref{eq:P1}, one takes the factor $1/\ell_{mn}$ out of the minimization,
 the same argument would lead to a  typical  re-organization size of  $\ell \simeq 1$ and to $P_1 - P_0 \propto L$. The numerical results obtained using our exact construction totally disagree with these predictions: $P_1 - P_0$ shrinks to zero with the system size and $\ell$ displays a beautiful scale-free behavior (Fig.~\ref{fig:central_results}.c and d).

To understand our results, it is important to study the statistics of $ \delta E_{mn} $.
 Its mean shows a growth compatible at large size with the already mentioned KPZ exponent, $\ell_{mn}^{\frac{1}{3}}$ (Inset Fig.~\ref{fig:gap_stats}).
   The gap distributions, rescaled by its mean, collapse more and more when $\ell_{mn} \to \infty$   
   to a universal curve  that vanishes at the origin with a power law similar to the level repulsion observed for the spectrum of random matrices.
However, it is crucial to note that the smallest gaps do not fall on this universal curve.
These rare, but almost vanishing, gaps are the ones  that determine $P_1$.
In particular, we observe the PDF of $ \delta E_{mn} $ (denoted by $\rho_{\ell}(\delta E)$) does not vanishes at the origin, but saturates at a value $\sim 1/\ell_{mn} $, as clearly shown by the plateau of Fig.~\ref{fig:gap_stats} \footnote{The SI shows that these rare degenerate ground states correspond to polymers that are spatially well separated. In this limit the repulsion between the two polymer configurations is very weak and very small gaps can be observed.}.
This observation is not only numerical, indeed it was shown that the probability to find almost degenerate  self-avoiding ground states is inversely proportional to the polymer length \cite{de2015crossing,de2015crossing,de2016crossing}.
 

We remark that Eq.~\eqref{eq:P1} requires the minimization of the energy cost per unit length. As this poses an additional difficulty, we perform the minimization in two steps. 
In the first step we consider all pairs of node $(m,n)$ with a given distance $\ell_{mn} =\ell$ and select the excitation with the minimal cost among them: $\delta e_{\ell} = \min\limits_{\ell_{mn}=\ell} \delta E_{mn}$.
We take the minimum among $L-\ell$ identically distributed random variables that are strongly correlated (as consecutive segments have a large overlap). The effective number of independent variables is $N_{\ell}=L/\ell$, which allows us to perform the statistics of the minimum among $N_{\ell}$ independent and identically distributed random variables. Given that $N_{\ell}$ is large and $\rho_{\ell}(\delta E\rightarrow 0 ) \sim 1/\ell$, we have 
\begin{eqnarray}
	&\text{Proba} [\delta e_{\ell}>x] = \left[1-\int_{0}^{x}\rho_{\ell}(\delta E)\right]^{N_{\ell}} \nonumber \\
	&\approx \exp\left(-N_{\ell}\int_{0}^{x}\rho_{\ell}(\delta E)\right) \approx \exp\left(-\frac{N_{\ell}}{\ell}x\right)
\end{eqnarray}
The second step takes into account all lengths to minimize the energy cost per length.
\begin{equation}
\label{eq:omegadef}
\frac{\Delta P_1}{L} = \min_{\ell} \frac{\delta e_{\ell}}{\ell}  
\end{equation}
We call $\omega_{\ell} = \delta e_{\ell}/\ell$ in Eq.~\eqref{eq:omegadef} and obtain  
\begin{equation}\label{eq:p_omega_ell}
p_{\ell}(\omega_{\ell}) \approx N_{\ell}\exp\left(-N_{\ell}\omega_{\ell} \right )
\end{equation}
Thus, the gap $\Delta P_1/L$ is also exponentially distributed with a mean $\overline{\Delta P}_1 /L= \left( \sum_{\ell} N_{\ell} \right)^{-1}\sim 1/\left(L\log(L/2)\right)$, which is consistent with the behavior shown by the red curve in Fig.~\ref{fig:central_results}.d . Moreover, discarding the factor $1/\ell_{mn}$ in the minimization of Eq.~\eqref{eq:omegadef}, one recovers the expected exponential distribution with a mean $\left(\sum_{\ell}\frac{N_{\ell}}{\ell} \right)^{-1}\sim L^{-1}$ (the blue curve in Fig.~\ref{fig:central_results}.d).

Finally we compute the statistics of the size of the second path $P_{\ell}$:
\begin{eqnarray}
P_{\ell} &=& \int d \omega_{\ell}   p_{\ell}(\omega_{\ell}) \prod_{\ell' \ne \ell} \int_{\omega_{\ell}}^{\infty} p_{\ell'}(\omega_{\ell'}) \sim \ell^{-1}
\end{eqnarray}
which is coherent with our observations for all types of disorders shown by the red curves in Fig.~\ref{fig:central_results}.c. Similarly if one discards the factor $1/\ell$ in the minimization, one obtains the length distribution scales as $\ell^{-2}$ as shown by the blue curves in Fig.~\ref{fig:central_results}.c.

\paragraph{Conclusions}

In this paper, we have studied the nonlinearity of the Darcy's law for a yield stress fluid in a 2D porous material. We have shown that the onset of flow is associated  to a plastic depinning transition. In this context, power law  behaviors have been found \cite{Aussillous201609023} but here, we clearly identify a divergent scale in the length of the new channels.
 Our results are idependent of the disorder distribution suggesting a robust universality and can be easily extended in 3D.

It would be interesting to study the fractal properties of the flowing region.
For example, in \cite{chevalier15a,chevalier15c} the statistics of the void was characterized numerically.
Moreover, exact results have been obtained for the geometry of the branching paths of the  first passage percolation model relevant for delta rivers \cite{barraquand2018} and animal trails \cite{kawagoe2017}.

\paragraph{Acknowledgements. ---}
This work is supported by "Investissement d'Avenir" LabEx PALM (ANR-
10-LABX-0039-PALM) and by the European Union's Horizon 2020 research and innovation programme under the Marie Sklodowska-Curie Grant Agreement No.
794750 (A.D.L.).

\appendix 


%

\end{document}


\title{Darcy law for yield stress fluid}
\author{Chen Liu}
\affiliation{FAST, CNRS, Univ. Paris-Sud, Universit\'{e} Paris-Saclay, 91405 Orsay, France}
\author{Andrea De Luca}
\affiliation{LPTMS, CNRS, Univ. Paris-Sud, Universit\'{e} Paris-Saclay, 91405 Orsay, France}
\author{Alberto Rosso}
\affiliation{LPTMS, CNRS, Univ. Paris-Sud, Universit\'{e} Paris-Saclay, 91405 Orsay, France}
\author{Laurent Talon}
\affiliation{FAST, CNRS, Univ. Paris-Sud, Universit\'{e} Paris-Saclay, 91405 Orsay, France}

\begin{abstract}
In this supplementary information, we present some details of our computing methods and additional figures that complete the results presented in the main text. We use the same notations in the following as introduced in the main text. 
\end{abstract}

\maketitle
\section{Dependence of local pressure $p_i$ on the imposed pressure $P$ for a given flowing network $\mathcal{L}(P)$}
The local pressure of a site $i\in\{1,2,3,\ldots,N_c\}$  belonging to the flowing network $\mathcal{L}(P)$ writes:
\begin{equation} \label{eq:linear}
	p_i = a_iP+b_i \quad .
\end{equation}
We describe how to determine the two coefficients $a_i$ and $b_i$. 
We denote  $i=1$ for the inlet node and $i=N_c$ for the outlet node.
 The following calculations can be  easily generalized to more complicated boundary conditions with multiple inlet and outlet nodes. We define the two following matrices:
\begin{itemize}
	\item The  symmetric adjacent matrix $A$ associated to $\mathcal{L}(P)$. Namely, $A_{ij}=1$ when the throat $(ij)$ is flowing and $A_{ij}=0$ otherwise.
In particular, the diagonal elements are null.
	\item The  matrix $I$ that represents the direction of the flow through a throat $(ij)$ ($+1$ if the flow is in the direction $i$ to $j$ and $-1$ otherwise). In the directed path approximation, we assume that the flow is always flowing from the inlet to the outlet direction.
\end{itemize}
The local debits satisfy the linear relation
$q_{ij}=A_{ij}(p_{i}-p_{j})-I_{ij}\tau_{ij}$.
The Kirchoff condition then writes:
\begin{equation}\label{eq:kirchoff}
\sum _{j=1}^{N_c}A_{ij} (p_i - p_j)    = \sum _{j=1}^{N_c} I_{ij}\tau_{ij},
\end{equation}
together with  the pressure imposed at the inlet   $p_1= P$ and at the outlet $p_{N_c}= 0$.
This can be recast in the matrix form:
\begin{equation}\label{Exact}
	\mathbf{M}\mathbf{p} = \mathbf{u} + P\mathbf{v}    
\end{equation} 
where 
\begin{align}\label{eq:vu}
	\mathbf{p}& =\left(p_2, p_3, \ldots, p_{N_c-1} \right) \\
	\mathbf{u} &= \left(\sum _{j=1}^{N_c} I_{2j}\tau_{2j}, \; \sum _{j=1}^{N_c} I_{3j}\tau_{3j}, \ldots, \sum _{j=1}^{N_c} I_{N_c-1,j}\tau_{N_c-1,j} \right) \\
	\mathbf{v} &= \left(  A_{21}, A_{31},\ldots, A_{N_c-1,1}   \right)
\end{align}
$$\mathbf{M} = 
\begin{pmatrix}\label{matrix}
\sum _{j=1}^{N_c}A_{2j}& -A_{23} & \ldots & -A_{2,N_c-1} \\
-A_{32} & \sum _{j=1}^{N_c}A_{3j} & \ldots & -A_{3,N_c-1} \\
\ldots & \ldots & \ldots & \ldots \\
-A_{N_c-1,2} & -A_{N_c-1,3} & \ldots & \sum_{j=1}^{N_c}A_{N_c-1,j}
\end{pmatrix}.
$$		
Leading to the vector form of Eq.\eqref{eq:linear} with
\begin{eqnarray}\label{eq:coeff}
	\mathbf{a} &=& \mathbf{M}^{-1}\mathbf{v} \nonumber  \\ 
	\mathbf{b} &=& \mathbf{M}^{-1}\mathbf{u}
\end{eqnarray}
Note that the coefficients $\mathbf{a}$, which determine the permeability $\kappa$  depends only on the topology of $\mathcal{L}(P)$ through the adjacent matrix $A$. In abscence of local thresholds $\mathbf{u}=0$, we retrieve the Newtonian fluid problem. We conclude that the permeability of $\mathcal{L}(P)$ for the  yield stress fluid is the same for a Newtonian fluid. 

\section{Universal behaviors}
We tested three different distributions for  local thresholds  with mean  $1.8$ and standard deviation $0.2$:  uniform, Gaussian and exponential.

\begin{figure}
	\begin{center}
	\begin{minipage}{0.49\hsize}
		\includegraphics[width =\hsize,trim = 50 0 10 0, clip=false]{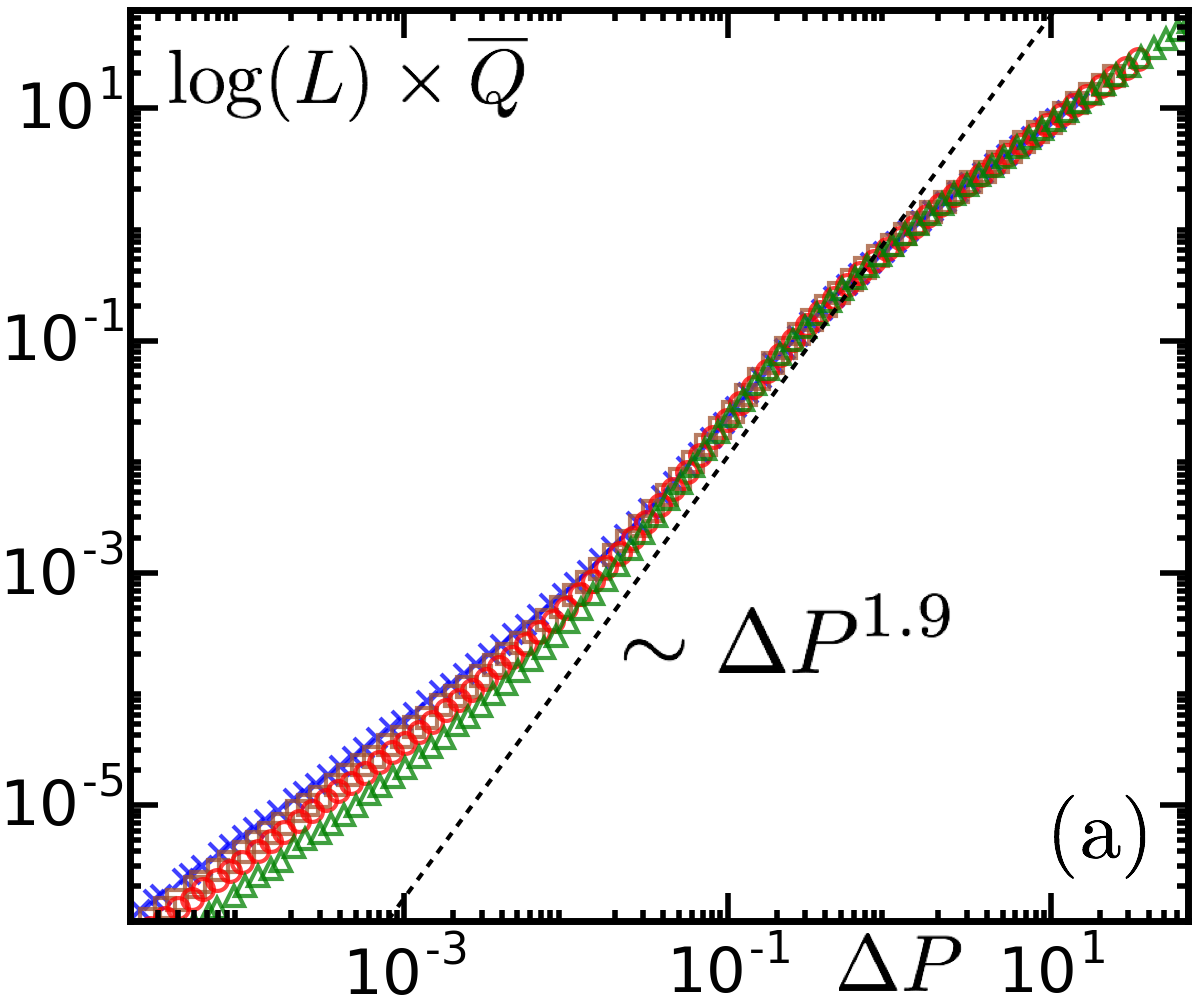}
	\end{minipage}
	\begin{minipage}{0.49\hsize}
		\includegraphics[width =\hsize,trim = 10 0 50 0, clip=false]{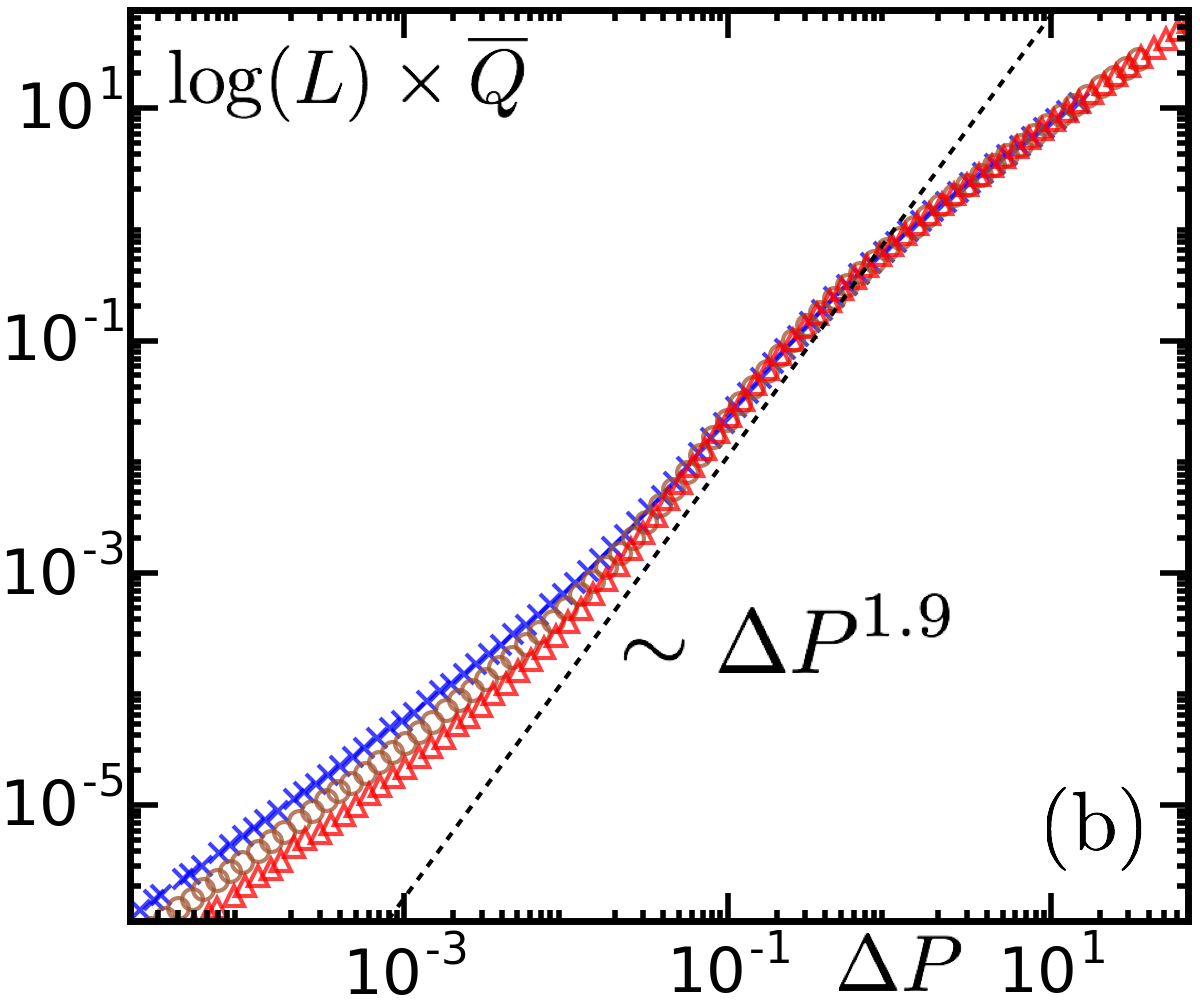}
	\end{minipage}
	\begin{minipage}{0.49\hsize}
	\includegraphics[width =\hsize,trim = 50 0 10 0, clip=false]{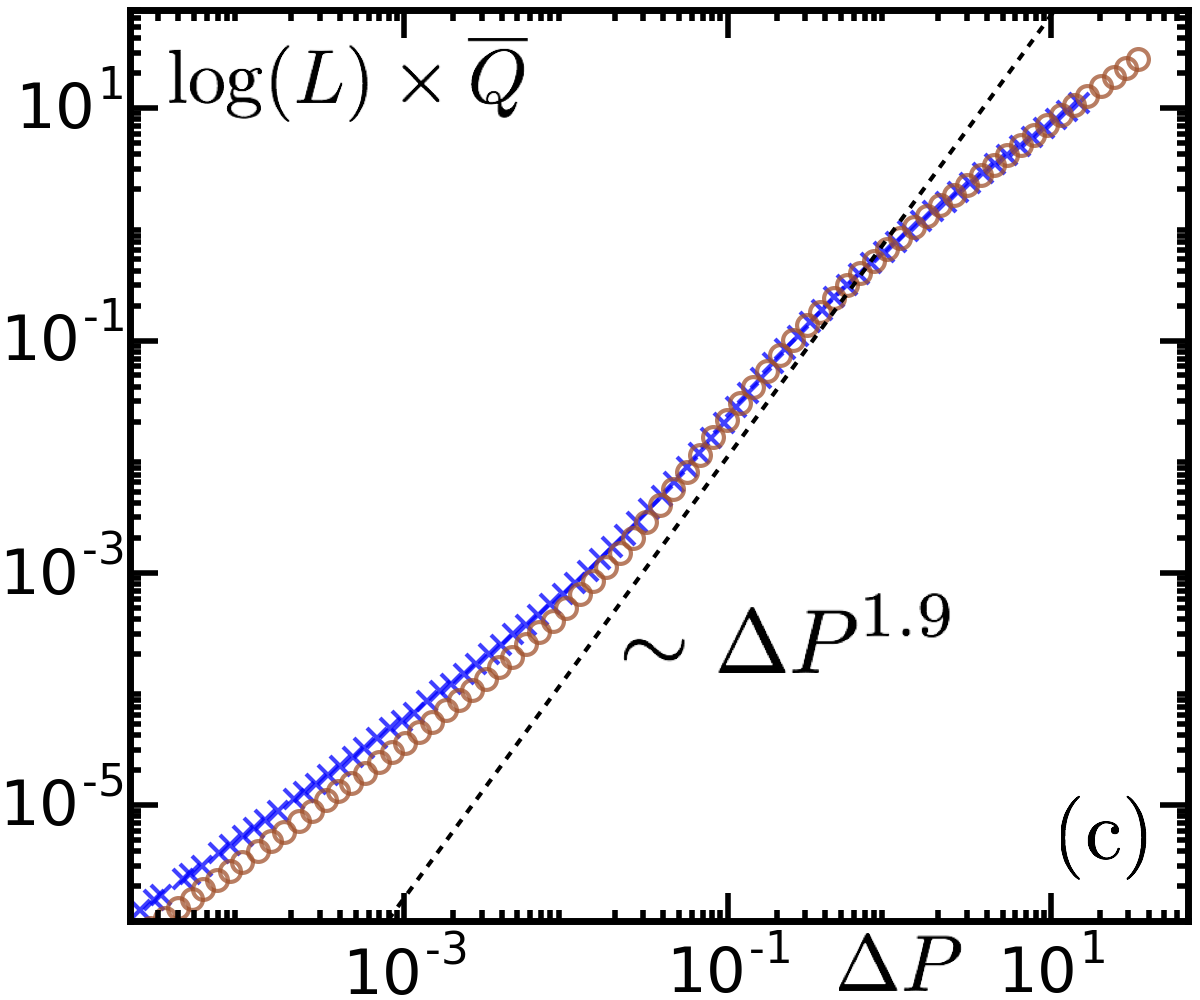}
	\end{minipage}
	\begin{minipage}{0.49\hsize}
	\includegraphics[width =\hsize,trim = 10 0 50 0, clip=false]{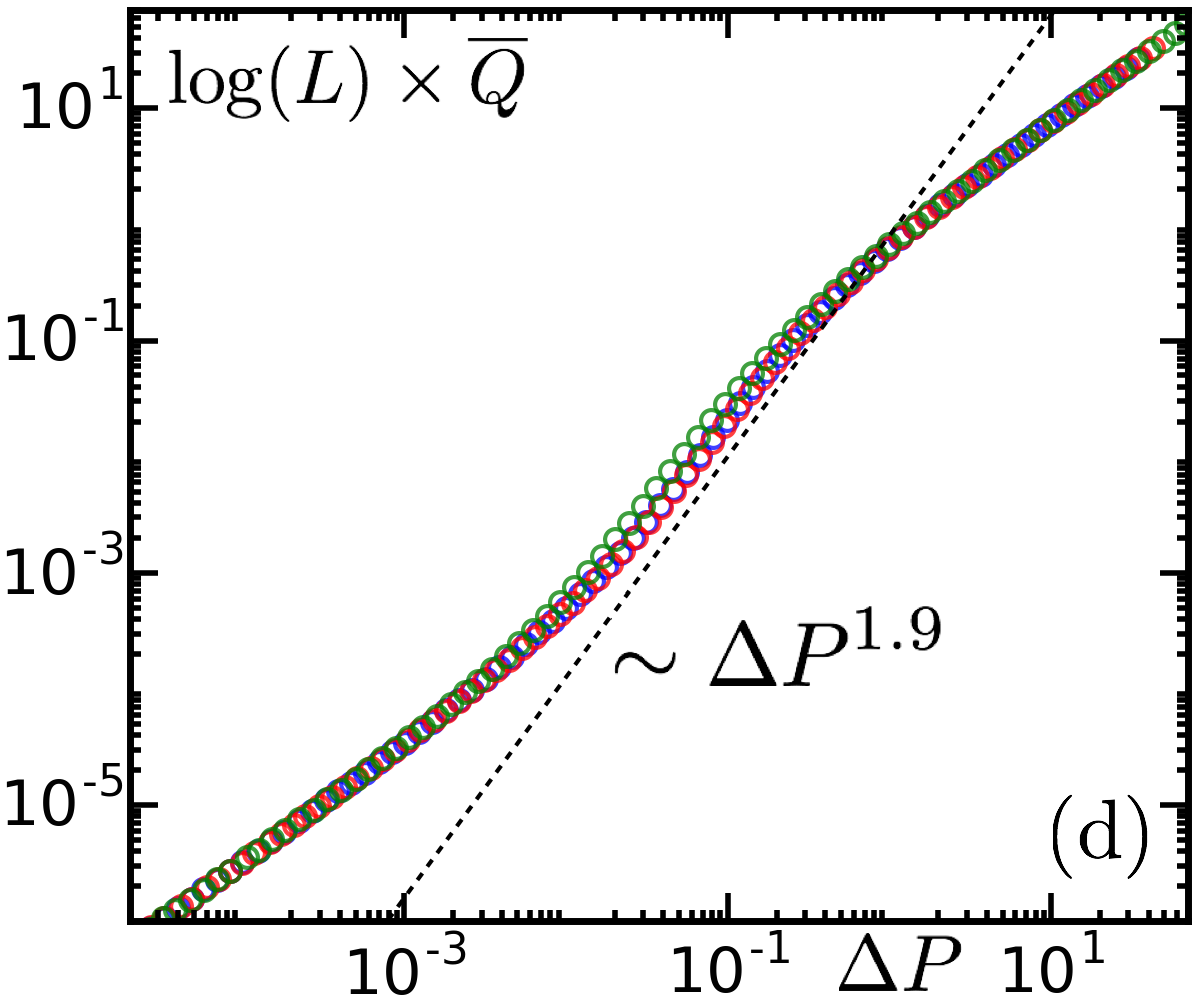}
\end{minipage}
\end{center}
	\caption{Rescaled mean flow curves $\log(L)\overline{Q}$ versus $P-P_0$ for different system sizes $L$, where $\overline{Q}$ has the same definition as Fig.3.a of the main text. (a), (b) and (c) correspond respectively to the types of disorder: uniform, Gaussian, exponential.  Different symbols for different system sizes: x: $L=64$, $\square$: $L=100$, $\bigcirc$: $L=128$, $\triangle$: $L=256$. The collapse of curves for different system sizes at very high pressure confirms the analytical prediction. (d) Data for different types of disorder (Blue: uniform; Red: Gaussian; Green: Exponential) for the same system size $L=128$ are plotted. }
	\label{fig:kappa_infity}  
\end{figure}

\paragraph{ Permeability at high pressure.} Analytical calculations show that at very high pressure, the permeability writes
\begin{equation}\label{eq:kappa_infity}
	\kappa_{\infty} \sim \frac{L}{\log(L)} \quad .
\end{equation}
This is verified by our numerical results for all types of disorder in Fig.\ref{fig:kappa_infity}.

\begin{figure}
	\begin{center}
		\begin{minipage}{0.49\hsize}
			\includegraphics[width =\hsize,trim = 50 0 10 0, clip=false]{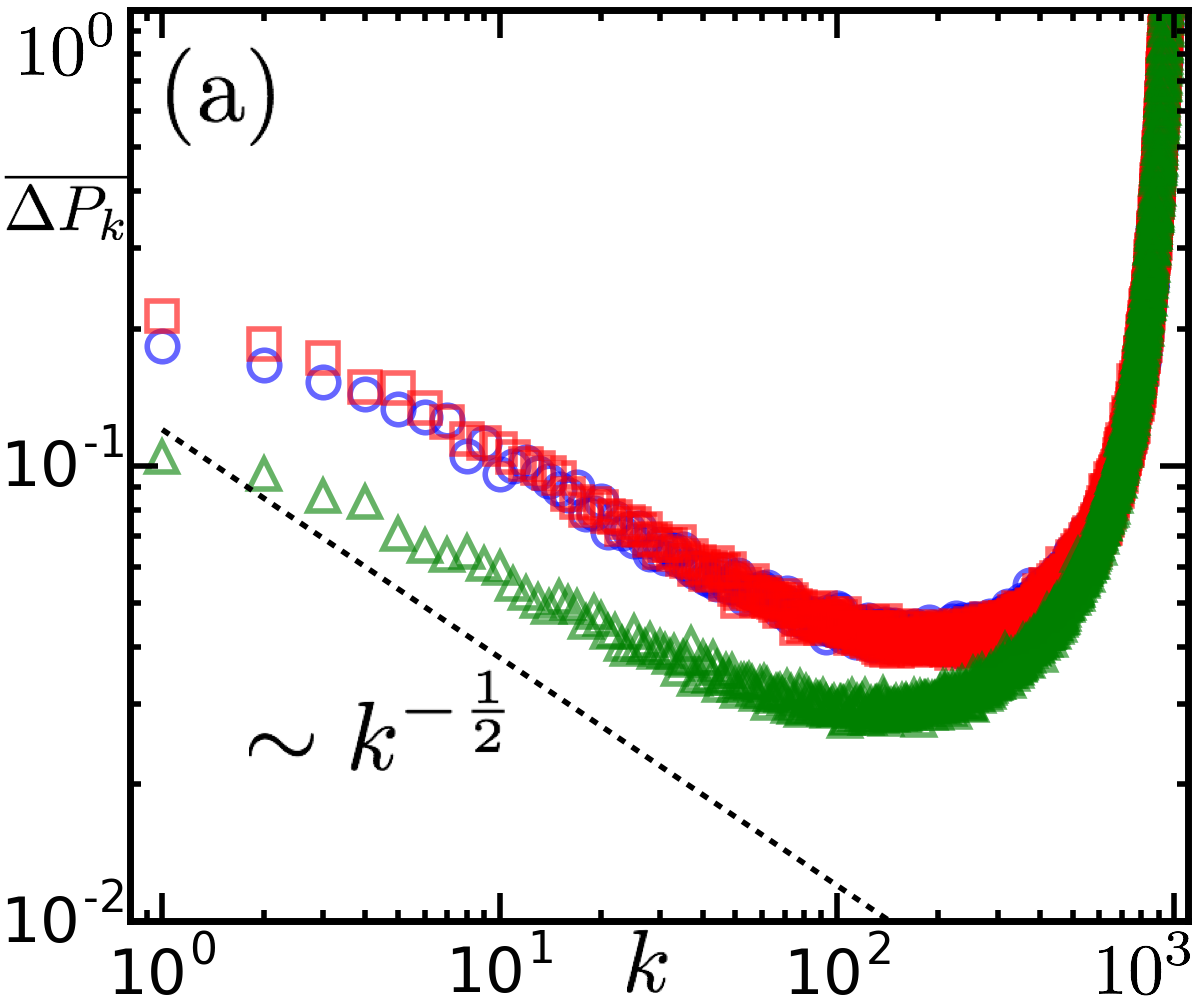}
		\end{minipage}
		\begin{minipage}{0.49\hsize}
			\includegraphics[width =\hsize,trim = 10 0 50 0, clip=false]{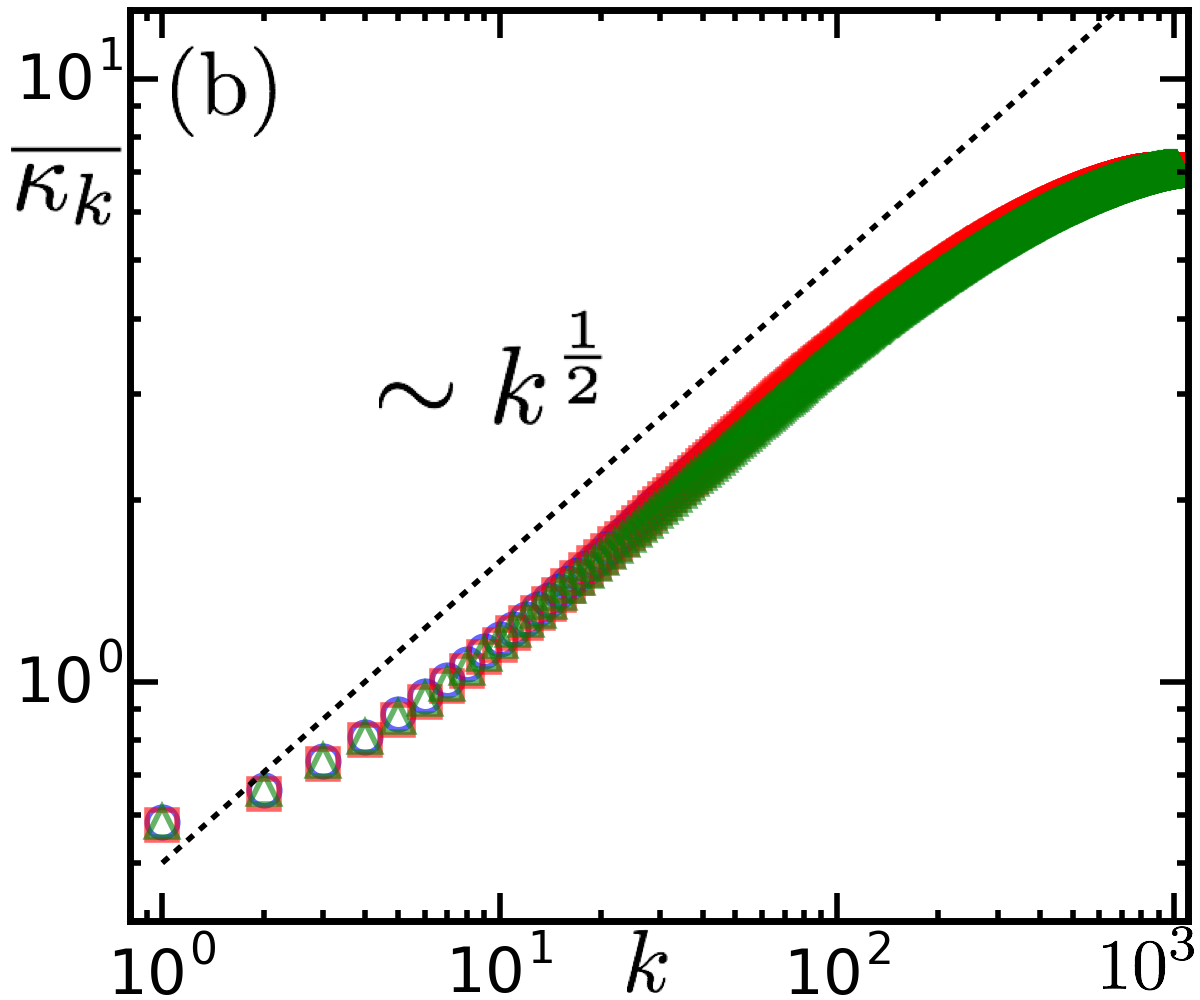}
		\end{minipage}
	\end{center}
	\caption{(a) Averaged pressure increments $\overline{\Delta P_k}$ and (b) averaged system effective permeability $\overline{\kappa_k}$ versus path number $k$ for three different types of disorder and for the same system size $L=64$. $\bigcirc$: uniform; $\square$: Gaussian; $\triangle$: exponential.}
	\label{fig:universal}   
\end{figure}

\paragraph{Scaling of the intermediate regime.} The mean flow curve for different types of disorder are shown in Fig.\ref{fig:kappa_infity}(d). The three curves display similar scaling exponent $\beta$,  consistent with a quadratic relationship between the flow rate $Q$ and the pressure  $\Delta P\equiv P-P_0$. 

Moreover, the scaling in $k$ of the mean permeability $\overline{\kappa_k}$ and the mean pressure increment $\overline{\Delta P_k}$ seems to be disorder independent (see  Fig.\ref{fig:universal}).
 This indicates that the flowing network $\mathcal{L}(P)$ evolves in a universal fashion with increasing $P$ despite of different types of disorder. 

\begin{figure}
	\begin{center}
		\begin{minipage}{0.49\hsize}
			\includegraphics[width =\hsize,trim = 50 0 10 0, clip=false]{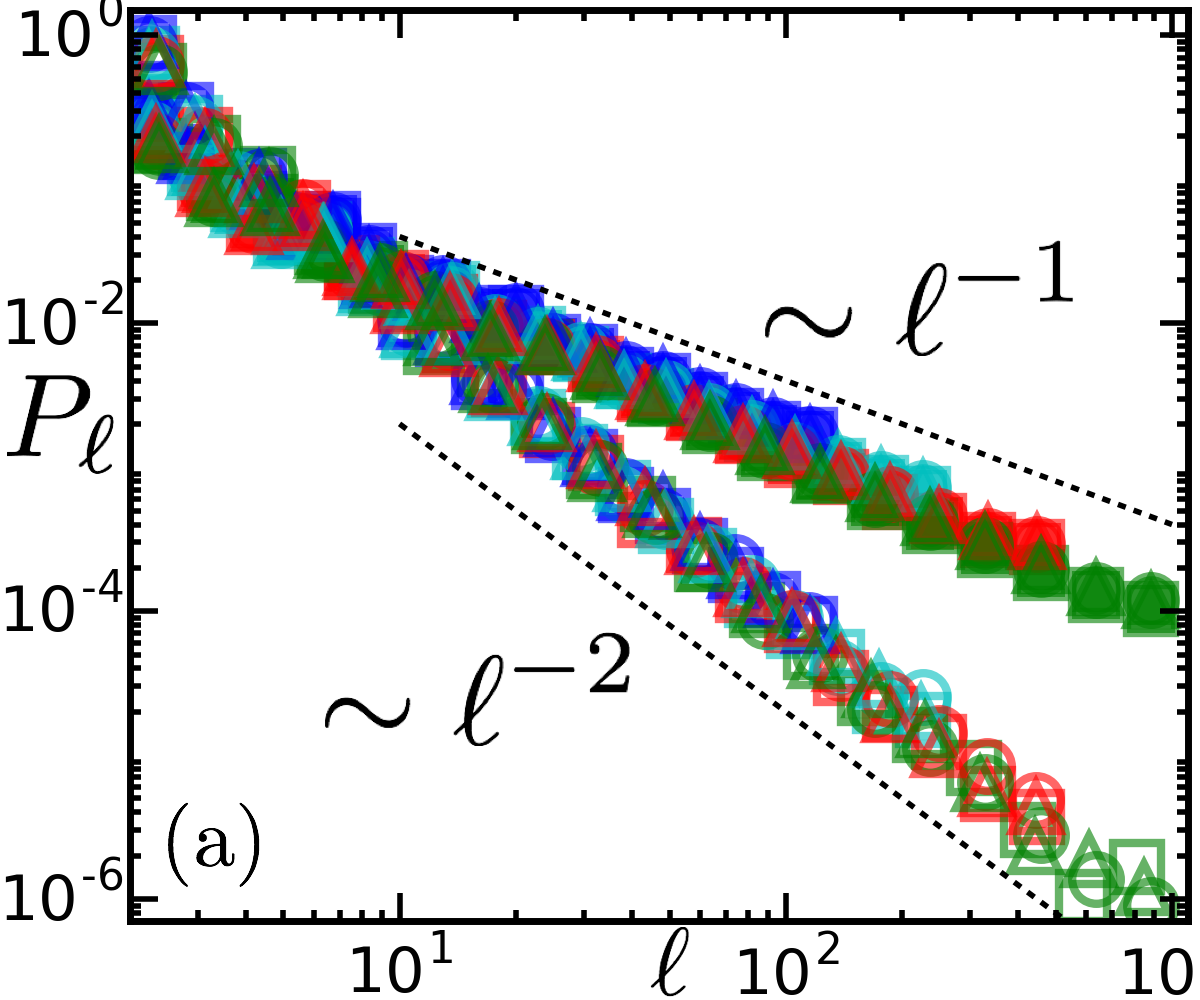}
		\end{minipage}
		\begin{minipage}{0.49\hsize}
			\includegraphics[width =\hsize,trim = 10 0 50 0, clip=false]{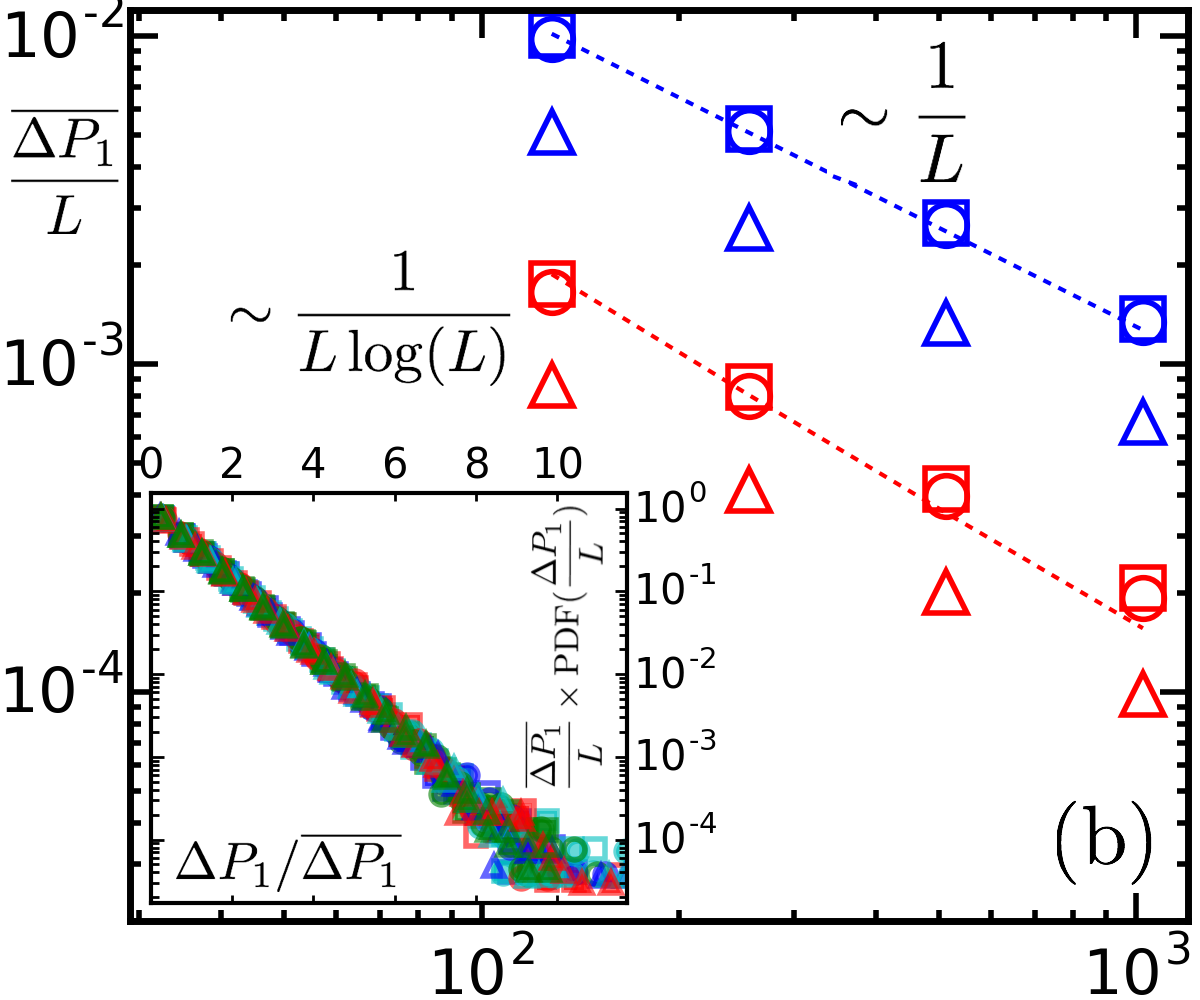}
		\end{minipage}
	\end{center}
	\caption{(a)-Distribution $P_{\ell}$ of the length of the second path $C_1$, (b)-Mean first gap $\overline{\Delta P_1}/L$  versus the system size $L$ and (b)inset-Renormalized PDF of the first gap for all system sizes (Blue: $L=128$; Cyan: $L=256$; Red: $L=512$; Green: $L=1024$) and all types of disorder ($\bigcirc$: uniform; $\square$: Gaussian; $\triangle$: exponential). In (a) and the inset of (b), empty symbols for the actual problem of porous medium and filled symbols for the results when neglecting the factor $1/\ell$ in the minimization in Eq.(11) in the main text.}
	\label{fig:universal_c1}  
\end{figure}

\paragraph{Universal statistics of the second excited path.  }
Similarly, as shown in Fig.\ref{fig:universal_c1}, the statistics of the second flowing path is independent of the type of disorder and display a robust universality.

\section {Characteristics of the  ground and the first excited state with self-avoiding condition}
 In this section, we take a closer look at the ground state configuration and at the first excited state for different values of the energy gap $\delta E$. 

\begin{figure}
	\begin{center}
	\begin{minipage}{0.46\hsize}
		\includegraphics[width =1.1\hsize,trim = 50 0 10 0, clip=false]{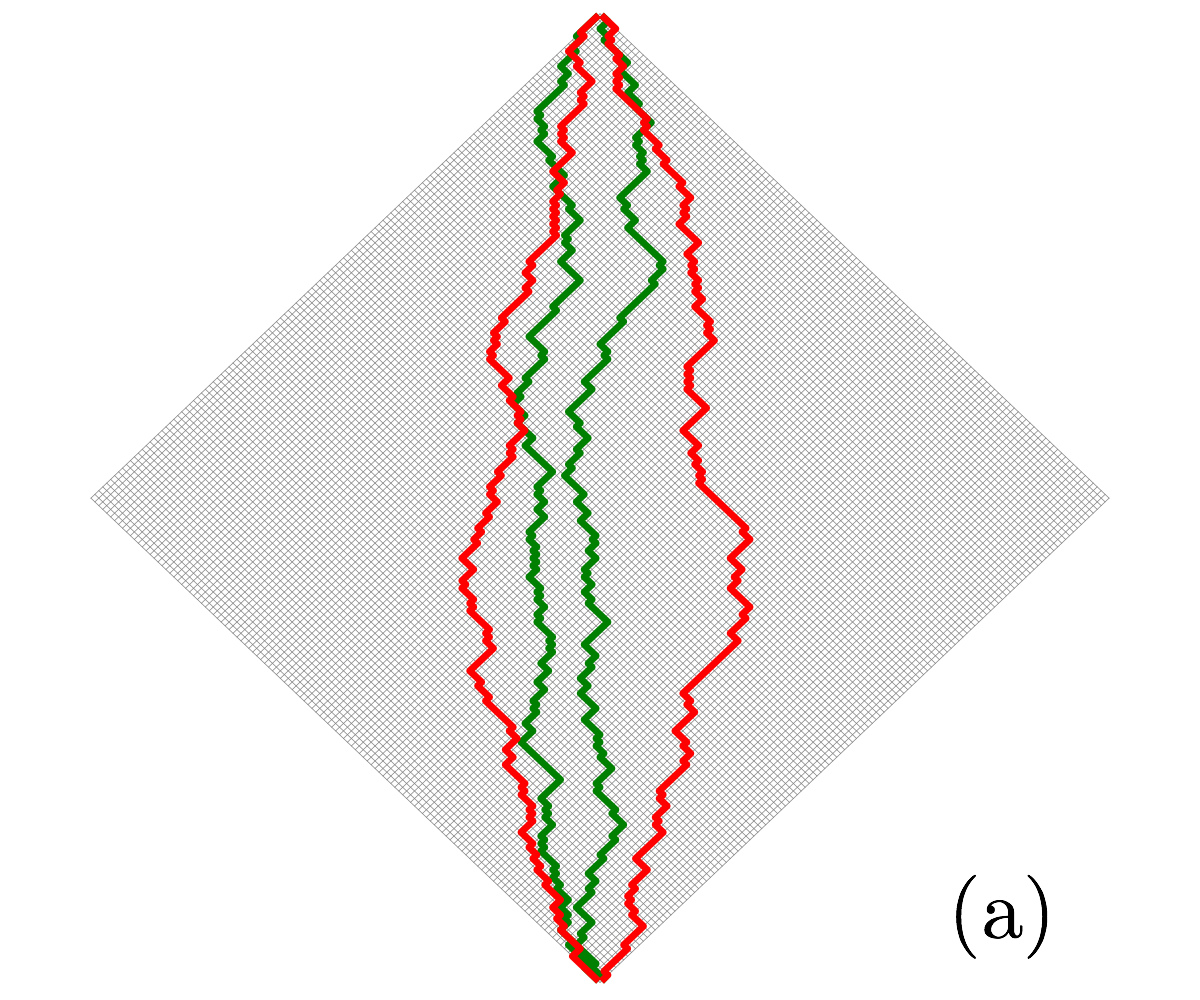}
	\end{minipage}
	\begin{minipage}{0.49\hsize}
		\includegraphics[width =0.95\hsize,trim = 10 0 50 0, clip=false]{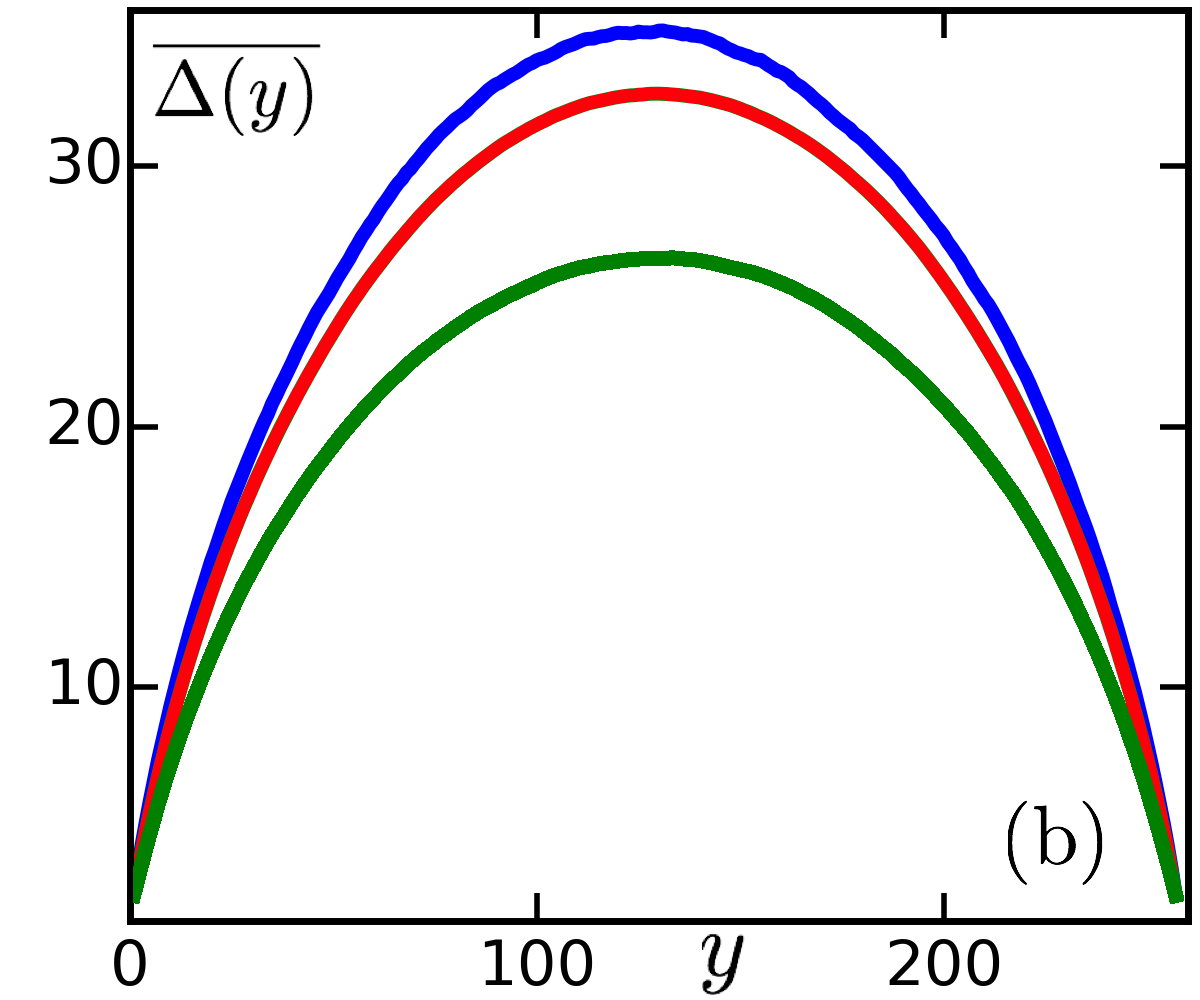}
	\end{minipage}
\end{center}
\caption{(a) Typical configurations of the two first self-avoiding ground states for small energy gap $\delta E \approx 0.069$ (red) and large energy gap $\delta E \approx 5.3$ (green). (b) Averaged difference profiles for different values of energy gaps: blue: $\delta E \approx 0.1$; green: $\delta E \approx 1$, red: $\delta E \approx 10$} 
\label{fig:profile}
\end{figure}

When the two states are quasi-degenerated, i.e. $\delta E$ very small, one expects that the two directed polymers explore very different areas of the random medium.
  Conversely, if they are close to each other, a strong level repulsion should be expected.
This conjecture is confirmed by  typical configurations found for a small and a large value of $\delta E$ and by the results obtained by  averaging over large amount of realizations (see Fig.\ref{fig:profile}).

In particular, for Fig.\ref{fig:profile}).b, we have defined the difference $\Delta(y)$ between the two paths as 
\begin{equation}
\Delta (y) = |x(y)-x'(y)| \quad .
\end{equation}
We compute the averaged difference profile $\overline{\Delta (y)}^{\delta E}$ for different values of $\delta E$. The corresponding averaged profiles clearly show that the difference profile becomes smaller when increasing the energy gap. 

\bibliography{biblio}